\documentclass[11pt]{article}
\usepackage{amsmath}
\usepackage{amssymb}
\usepackage{graphicx}  % Include figure files
\usepackage[a4paper]{geometry}
\usepackage{authblk}
\usepackage{cite}

\graphicspath{{figures/}}

\newcommand{\email}[1]{\footnote{#1}}
\newcommand{\affiliation}[1]{\small\textit{#1}\\}

\newcommand{\Sussex}{\affiliation{
Department of Physics and Astronomy,
University of Sussex, Falmer, Brighton BN1 9QH,
U.K.}}

\newcommand{\HIPetc}{\affiliation{
Department of Physics and Helsinki Institute of Physics,
PL 64, 
FI-00014 University of Helsinki,
Finland
}}

\newcommand{\EHU}{\affiliation{
Department of Physics,
University of the Basque Country UPV/EHU, 
48080 Bilbao,
Spain
}}
\usepackage{xcolor}

\newcommand{\Tufts}{\affiliation{
Institute of Cosmology, Department of Physics and Astronomy, 
Tufts University,
Medford, MA 02155,
USA}}

\title{Comment on ``More Axions from Strings''}

\author[1,2]{Mark Hindmarsh\email{mark.hindmarsh@helsinki.fi}}
\author[3]{Joanes Lizarraga\email{joanes.lizarraga@ehu.eus} }
\author[4]{Asier Lopez-Eiguren\email{asier.lopez\_eiguren@tufts.edu}}
\author[3]{Jon Urrestilla\email{jon.urrestilla@ehu.eus}}
\affil[1]{\HIPetc}
\affil[2]{\Sussex}
\affil[3]{\EHU}
\affil[4]{\Tufts}
\date{\today}

\begin{document}

\maketitle

\begin{abstract}
We comment on a claim that axion strings show a long-term logarithmic increase in the number of Hubble lengths per Hubble volume \cite{Gorghetto:2020qws}, thereby violating the standard ``scaling'' expectation of an O(1) constant. We demonstrate that the string density data presented in Ref.~\cite{Gorghetto:2020qws} are consistent with standard scaling, at a string density consistent with that obtained by us \cite{Hindmarsh:2019csc,Hindmarsh:2021vih} and other groups. A transient slow growth in Hubble lengths per Hubble volume towards its constant scaling value is explained by standard network modelling \cite{Hindmarsh:2021vih}. 
\end{abstract}

The paper \cite{Gorghetto:2020qws} reports on and interprets the results of a set of numerical simulations of axion string 
networks in a complex U(1) field model of the axion, aiming to pin down the axion number density in the post-inflationary PQ symmetry-breaking 
scenario, and thereby provide an accurate prediction of the axion mass for dark matter searches. 

A central claim of the paper (first made in Ref.~\cite{Gorghetto:2018myk}) 
is that the long-established picture of scaling in string networks 
 \cite{Hindmarsh:1994re,Vilenkin:2000jqa}
does not apply to global strings, and that the average number of Hubble lengths of string per Hubble volume (proportional to $\xi$ in 
their notation) grows logarithmically with cosmic time in the long term, rather than tending to an O(1) constant.  If confirmed, this would have important implications for axion dark matter in the axion string scenario: in particular, the axion mass estimate would be significantly changed. 
Resonant cavity axion detectors \cite{Braine:2019fqb} benefit greatly from accurate mass estimates in order to reduce the search time.

Data from earlier simulations in the same field theory  
\cite{Yamaguchi:1998gx,Yamaguchi:1999yp,Yamaguchi:1999dy,Yamaguchi:2002sh,
Hiramatsu:2010yu,Hiramatsu:2012gg,Kawasaki:2014sqa,Lopez-Eiguren:2017dmc}
analysed in the framework of the standard scaling scenario were consistent with $\xi \simeq 1$. 
Following the appearance of Ref.~\cite{Gorghetto:2018myk}, other groups have 
also reported slow growth of $\xi$ in simulations 
\cite{Kawasaki:2018bzv,Vaquero:2018tib,Buschmann:2019icd,Klaer:2019fxc}. 
This is argued in Refs.~\cite{Gorghetto:2018myk,Gorghetto:2020qws} to be the true long-term 
behaviour of an axion string network, replacing the standard scaling model. 

In the standard scaling model, the string network evolves towards constant string density parameter,
which is easily 
understood in terms of the reduced or increased rate of loop production in 
under- or over-dense networks.  
The loops evaporate into axions and massive scalar modes \cite{Saurabh:2020pqe}.
This model is given a mathematical expression as the ``one-scale'' 
model \cite{Kibble:1984hp} and  the ``velocity-dependent one-scale" model (VOS) 
\cite{Martins:1996jp,Martins:2000cs}, which adjusts the loop production rate 
according to the root mean square velocity of the strings. 
The VOS model in its simplest form
has been checked against numerical simulations of gauge string networks \cite{Moore:2001px,Correia:2019bdl} and it
has recently been shown to give a good description of 
the approach to scaling in global string networks \cite{Hindmarsh:2021vih}.

On the other hand, in Ref.~\cite{Gorghetto:2020qws} it is not clear what the model is, beyond 
a hypothesis that the string density parameter grows logarithmically at late times.  
An argument is given for expecting logarithms based on the increase in the effective string tension and the
decoupling of the axion field in an idealised model of axion strings \cite{Dabholkar:1989ju}, 
which is the subject of ongoing discussion \cite{Dine:2020pds}, 
but no dynamical model is proposed. In particular, it is not made clear 
how the logarithmic decoupling of the axion field should affect the string density in this way. 
Others have tried adapting the VOS model with a time-dependent string mass per unit length 
\cite{Martins:2018dqg,Chang:2019mza,Chang:2021afa}, but this still results in a constant 
string density parameter at late times. The fit of the simplest VOS model to the 
global string network evolution close to the scaling fixed point 
is already very good \cite{Hindmarsh:2021vih}, 
so the adaptation is not motivated by the simulations themselves. 

The growth of $\xi$ observed in the numerical simulations of Ref.~\cite{Gorghetto:2020qws} are presented as strong evidence 
for the hypothesis that the growth is logarithmic in the long term. 
However, as shown in Refs.~\cite{Hindmarsh:2019csc,Hindmarsh:2021vih}, 
growth of $\xi$ in simulations can be understood in terms of the slow approach of the 
dynamical variable $\xi$ to a constant-$\xi$ scaling solution with $\xi_\infty =  1.19\pm0.20$.   
As we demonstrate below, the data presented in Ref.~\cite{Gorghetto:2020qws}, 
are consistent with $\xi \to \text{O}(1)$, and with the results of Refs.~\cite{Hindmarsh:2019csc,Hindmarsh:2021vih}, 
when properly analysed in a framework allowing for transients. 
Criticisms of the claims of long-term logarithmic growth made in 
Ref.~\cite{Hindmarsh:2019csc}, which was published before Ref.~\cite{Gorghetto:2020qws} appeared, 
are not addressed in Ref.~\cite{Gorghetto:2020qws}, and Ref.~\cite{Hindmarsh:2019csc} is not cited.
 
The important quantities under study are the string density parameter $\xi$, 
defined in terms of the total string length $\ell$ as 
\begin{equation}
\xi  = \ell t^2/ \mathcal{V},
\end{equation} 
where $t$ is cosmic time and $\mathcal{V}$ is the simulation volume, 
and the mean string separation 
\begin{equation}
L = \sqrt{\mathcal{V}/\ell}.
\end{equation} 

\begin{figure}[htb!]
    \centering
    \includegraphics[width=0.7\textwidth]{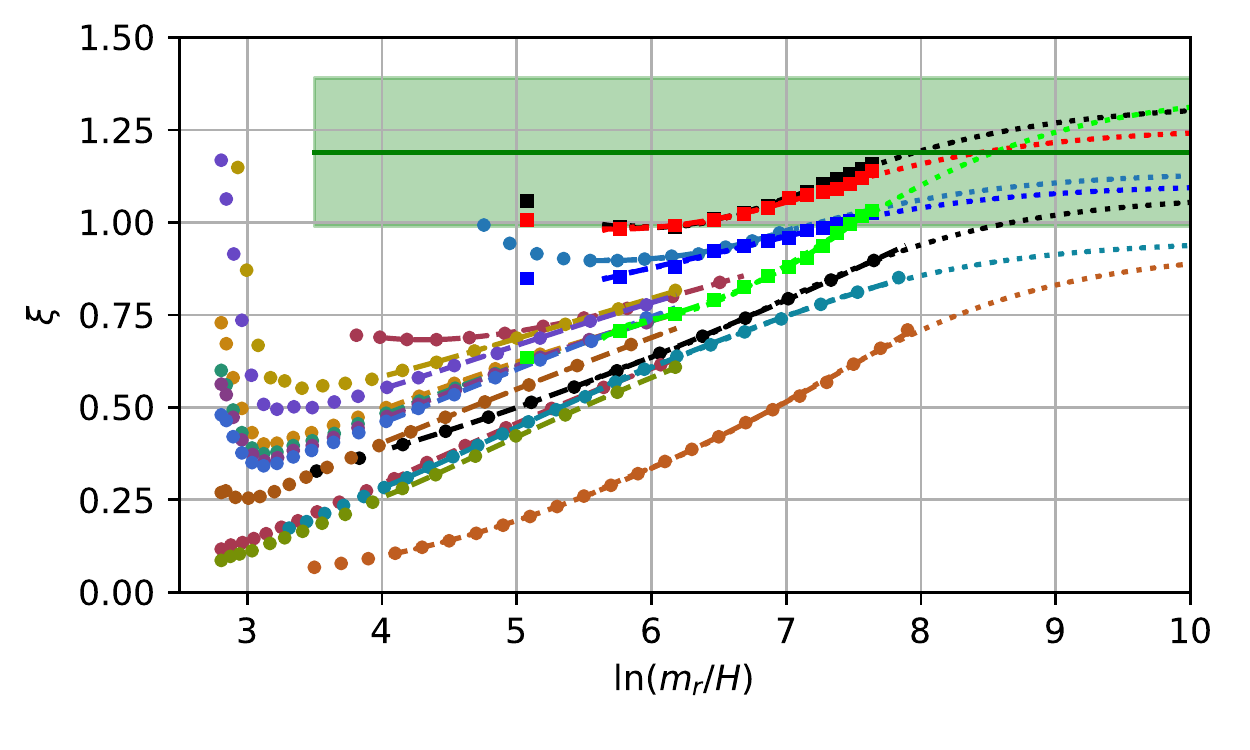}
    \caption{Data from Figure 1 of Ref.~\cite{Gorghetto:2020qws} (round markers) and \cite{Hindmarsh:2019csc} (square markers). 
 The dashed lines give fits to 
  (\ref{e:GHVeq}) with all parameters free, over ranges $\log(m_r/H) > 4$ or $\log(m_r/H) > 5.5$.      
    The dotted line shows the fit to a function (\ref{e:ss2}) with long-term linear growth, 
    as predicted by standard scaling, and an extrapolation. 
    The estimate of the asymptotic value of $\xi$ in \cite{Hindmarsh:2019csc} is shown as the green band.  
    All data are broadly consistent with an asymptotic value $\xi(t \to \infty) \simeq 1$.
         \label{fig:comp_log}}
 \end{figure}

To compare the results of Ref.~\cite{Gorghetto:2020qws} with ours, 
we have digitised data from Figure 1 of Ref.~\cite{Gorghetto:2020qws}
and present it together with data from \cite{Hindmarsh:2019csc} in Fig.~\ref{fig:comp_log}.
Data from Ref.~\cite{Gorghetto:2020qws} is presented  
as round dots  and data from \cite{Hindmarsh:2019csc} as square markers.  
Fig.~\ref{fig:comp_log} also shows 
the estimate of the asymptotic value of $\xi$ in \cite{Hindmarsh:2019csc} as a green band. 
Note that Ref.~\cite{Hindmarsh:2019csc} was more conservative about 
the preparation of the string networks, and waited longer before starting to record data. 
Note also that the choice made in Refs.~\cite{Gorghetto:2018myk,Gorghetto:2020qws}
to plot against the logarithm of cosmic time, rather than cosmic time, 
emphasises the early phase of the simulations, where the effect of initial conditions will be greater. 

One can see that the simulations of  \cite{Gorghetto:2020qws} and \cite{Hindmarsh:2019csc} 
give a consistent picture of the evolution of $\xi(t)$, and are distinguished only by the generally lower string density 
in Ref.~\cite{Gorghetto:2020qws}, resulting from the choice of initial conditions.
It is reassuring that the two data sets obtained from different codes, initial conditions, data collection methods, and number of simulations at each initial string density are in broad agreement. 
The different conclusions are therefore a result of the analysis rather than the simulations.

Eq.~(3) of Ref.~\cite{Gorghetto:2020qws} gives the authors' hypothesis for the 
behaviour of $\xi$, which we reproduce here 
\begin{equation}\label{e:GHVeq}
\xi = c_1 \log(m_r /H) + c_0 + \frac{c_{-1}}{\log(m_r /H)} + \frac{c_{-2}}{\log^2(m_r /H)}  ,
\end{equation}
where $m_r$ is the mass of the scalar field,  $H  = 1/2t$ is the radiation era Hubble rate, and 
$c_n$ are fit parameters. 

In Ref.~\cite{Gorghetto:2020qws} the data from all runs are fitted with  the first two terms ($c_1$ and $c_0$) 
as global (``universal'') fit parameters, and the value $c_{1} = 0.24(2)$ is presented. 
In the long-term logarithmic growth model, this would describe the network at large times. 
However, the ability to find a precise value for a global fit parameter does not in itself make the parameter physical: 
as the authors themselves point out, 
fits with a global log$^2$ term also give good results.  

\begin{figure}[htb!]
    \centering
    \includegraphics[width=0.5\textwidth]{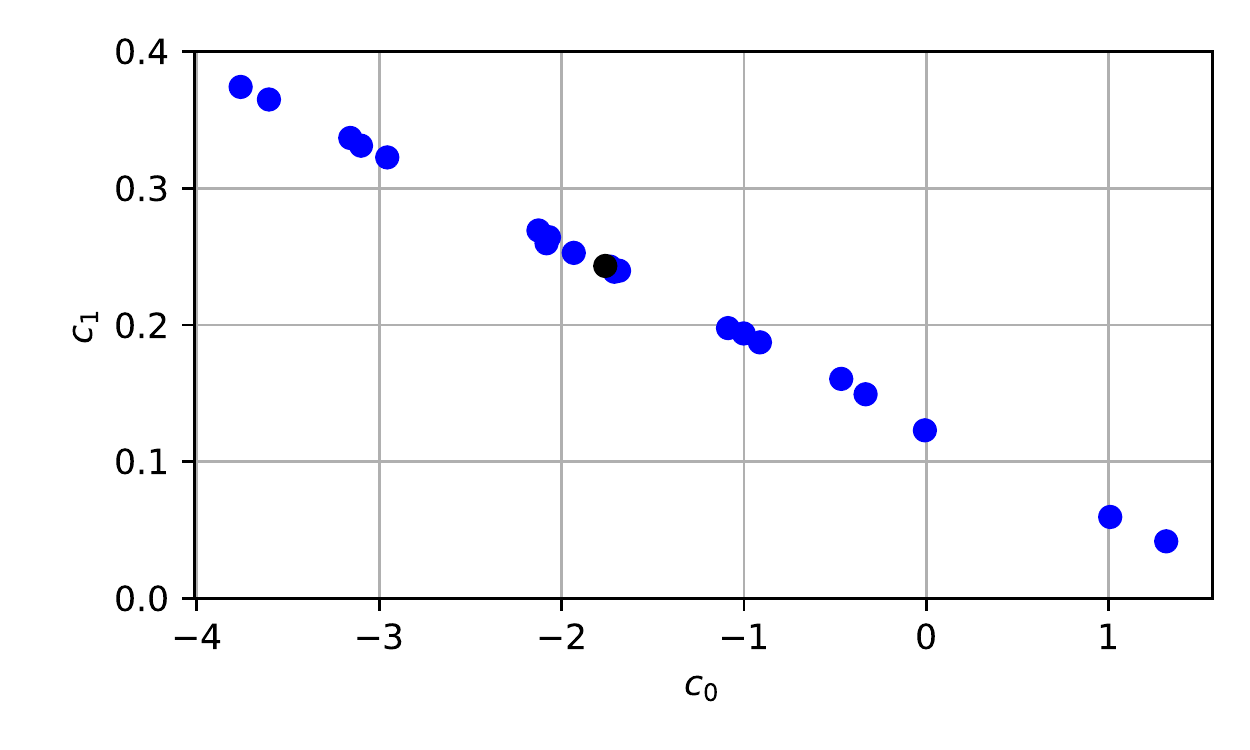}
    \includegraphics[width=0.5\textwidth]{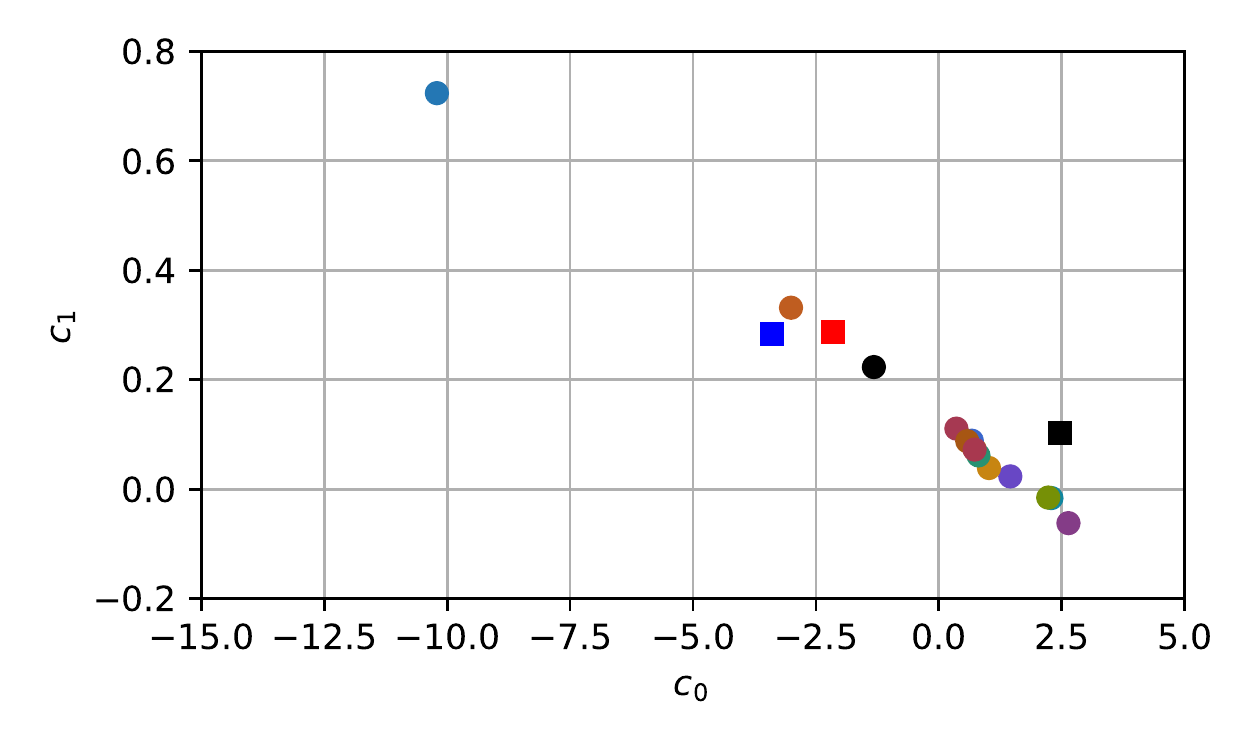}
    \caption{Top: Global fit parameters $c_1$ and $c_0$ of Eq.~\ref{e:GHVeq} fitted to the 14 digitised datasets from Ref.~\cite{Gorghetto:2020qws} (black),
along with 20 bootstraps of 14 samples (with replacement) of the same datasets (blue).  Each dataset is fitted with its own parameters $c_{-1}$ and $c_{-2}$. 
    Bottom: data fitted with four independent parameters $c_{n}$ for each dataset. 
    Datasets from Ref.~\cite{Hindmarsh:2019csc} are included. Markers have the same colour and shape code as Fig.~\ref{fig:comp_log}. 
    The dataset marked with the green square has $(c_0,c_1) = (-30, 1.8)$ and is outside the plot area. 
        \label{fig:comp_fit_log}}
 \end{figure}

If there is a physical parameter describing the long-term behaviour of the string network, 
it should be independent of the initial state. In order to check this independence, 
we have performed a set of 20 bootstrap fits with $c_1$ and $c_0$ as global fit parameters, 
and each dataset assigned its own $c_{-1}$ and $c_{-2}$. 
The bootstrap used samples of 14 (with replacement) from the 14 digitised datasets from \cite{Gorghetto:2020qws}. 
For each 30-parameter fit, the fit ranges were the same as \cite{Gorghetto:2020qws}:
$\log(m_r/H) > 4$, with one curve using  $\log(m_r/H) > 5.5$. 
The results for the global fit parameters $(c_0,c_1)$ are shown in the top panel of Fig.~\ref{fig:comp_fit_log}, with the result for the fit to all of the original 14 datasets shown in black.  For this fit we obtain $c_1 = 0.24$, which compares well with the quoted value $c_1 = 0.24(2)$ \cite{Gorghetto:2020qws}.  

The bootstraps show a wide variation in $c_1$, very strongly anticorrelated with $c_0$, showing that the uncertainty on $c_1$ arising from 
the choice of initial string density is much higher than the bracketed number would suggest.

Another way of seeing this sensitivity to the initial string density is to fit all four parameters $c_n$ individually to each dataset. 
The results of these fits, over the same ranges, is shown in the bottom panel of Fig.~\ref{fig:comp_fit_log}, 
along with the fits to the data from \cite{Hindmarsh:2019csc}, fitted over $\log(m_r/H) > 5.5$. 
One immediately notices that they also vary widely with initial string density, and are also strongly correlated. 

The uncertainties arising from the wide variation in the values of $c_1$ 
shows the logarithmic growth model has little utility in predicting 
the string density in the long term. 
This casts doubt that the parameter $c_1$ has any physical meaning.  

Let us now examine the consistency of fits to the data with standard scaling, 
which we recall is underpinned by the VOS model. 
The VOS model describes the network with two parameters, 
repre\-sen\-ting the efficiency with which loops are removed from long strings, and 
the efficiency with which mean curvature produces average acceleration. 
It predicts that the mean string separation $L = t / \sqrt{\xi}$ should grow 
linearly with cosmic time at late times, while the root mean square velocity tends to a constant.

In Ref.~\cite{Hindmarsh:2019csc} estimates for the asymptotic linear growth rate were extracted from the data by fits to 
\begin{equation}\label{e:ss}
L = x_* t + L_0 .
\end{equation}  
In standard scaling $x_*$ is predicted to be a constant O(1) physical parameter.
The phenomenological fit parameter $L_0$ is introduced 
to reduce the effect on estimates of $x_*$ of the initial conditions and the evolution of the RMS velocity. 
We obtained $x_* = 0.88 \pm 0.07$ for our largest simulations. 
We also looked for a slow increase in the value of $x_*$ in the data, finding none \cite{Hindmarsh:2019csc}.

The error budget included an exploration of the effect of different fit ranges,
which were taken in the half of the conformal time range of the simulations. 
Starting the fit too early biases the estimates of the asymptotic behaviour with 
transients associated with the evolution away from the initial state. 
These biases can be reduced by including velocity data and fitting with the VOS model, 
which accounts for some of this evolution \cite{Hindmarsh:2021vih}.

\begin{figure}[ht!]
    \centering
    \includegraphics[width=0.7\textwidth]{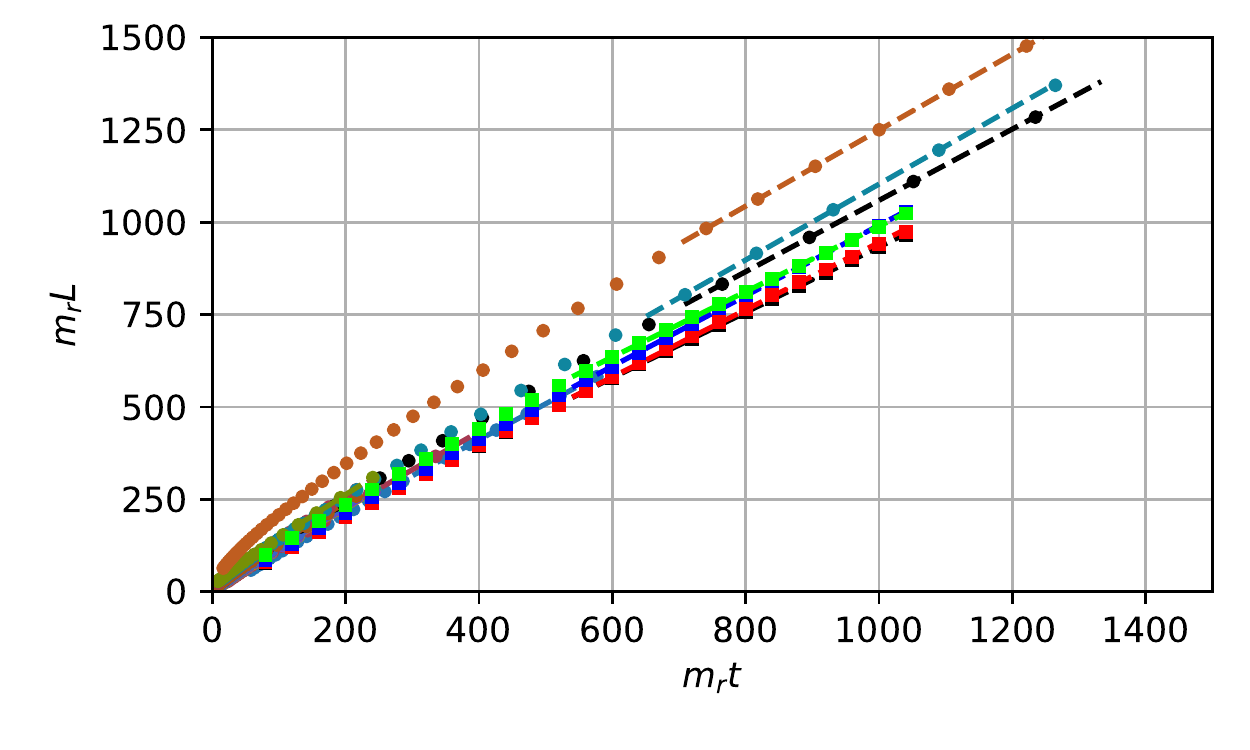}
    \caption{Data from Figure 1 of Ref.~\cite{Gorghetto:2020qws} (dots), and from \cite{Hindmarsh:2019csc} (square markers), 
   showing the mean string separation $L$. Dashed lines give fits to a straight line (\ref{e:ss2}) over the last half of the simulation 
   from which the asymptotic coefficient linear growth is extracted. 
    \label{fig:lin}}
 \end{figure}

We plot $L$ against $t$ in Fig.~\ref{fig:lin}, also with fits to (\ref{e:ss}) in the last half of the simulation, 
where the effect of initial conditions should be reduced. 
Again, data from Ref.~\cite{Gorghetto:2020qws} is plotted as dots, with data from 
\cite{Hindmarsh:2019csc} is plotted with square markers. 

\begin{figure}[ht!]
    \centering
    \includegraphics[width=0.5\textwidth]{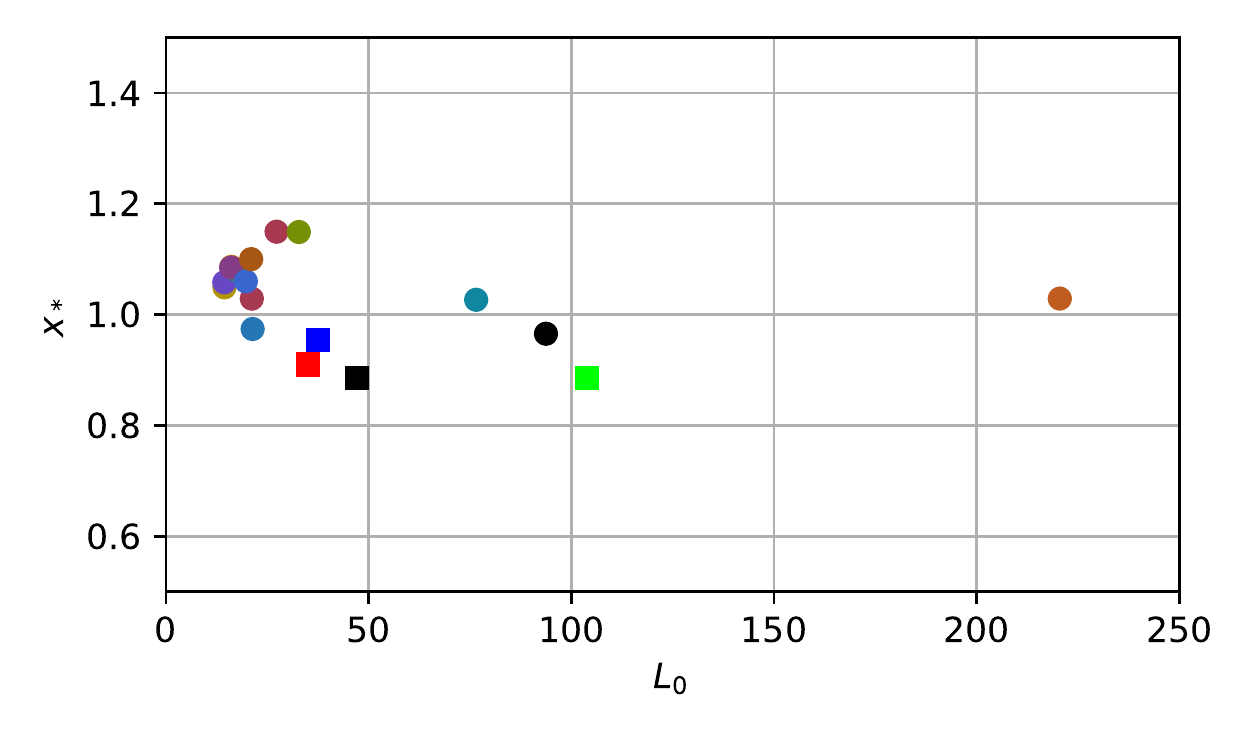}    
     \includegraphics[width=0.5\textwidth]{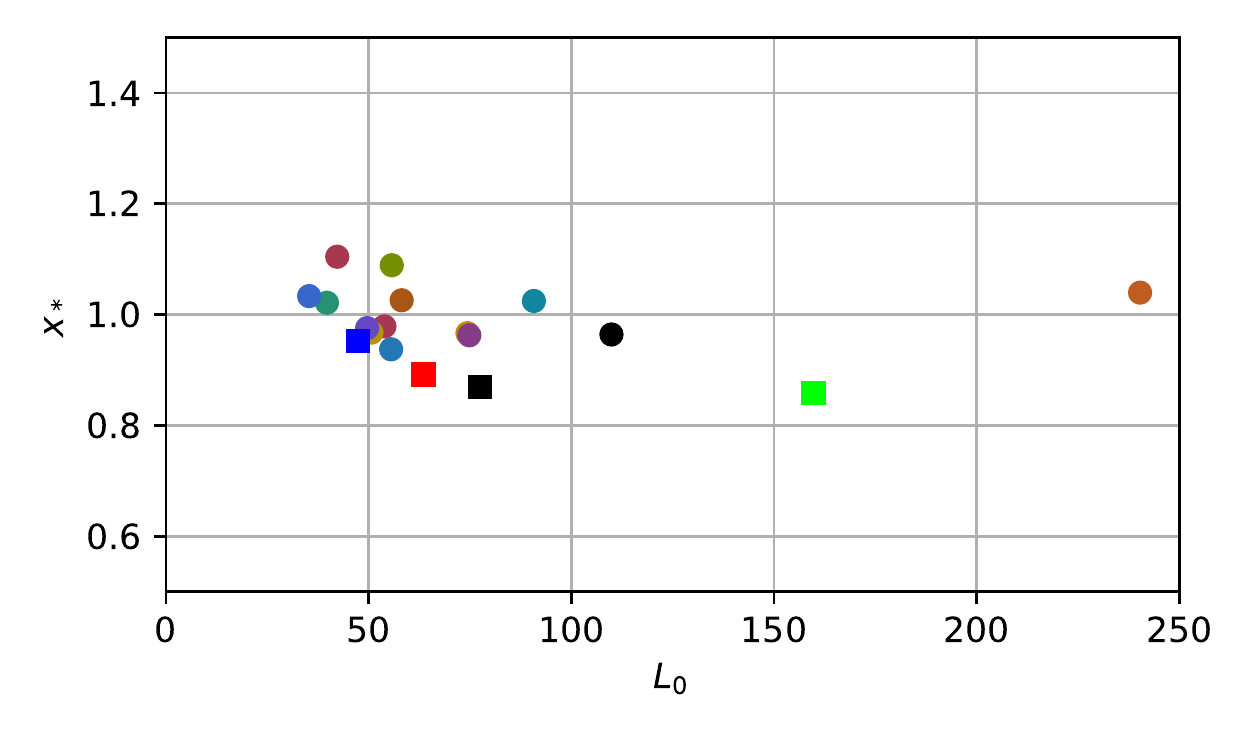}  
    \caption{Fit paramerers to the data from Figure 1 of Ref.~\cite{Gorghetto:2020qws} (dots), and from \cite{Hindmarsh:2019csc} (square markers). 
    Top: fit parameters $x_*$ and $L_0$ from the 2-parameter fit (\ref{e:ss}).
    Bottom: fit parameters $x_*$ and $L_0$ from the extended 4-parameter fit (\ref{e:ss2}).
    \label{fig:comp_ss}}
 \end{figure}

Fig.~\ref{fig:comp_ss} shows the best fit value of the parameters $x_*$ and $L_0$, and it is clear that all simulations are consistent with $x_* \simeq 1$, as predicted by standard scaling. There is no obvious correlation between the phenomenological fit parameter $L_0$ and the proposed universal one $x_*$,
indicating that the effect of different initial string densities is mostly captured by $L_0$.  

As a further comparison with the four parameter model of 
Eq.~(\ref{e:GHVeq}), we also fitted the data to a 4 parameter model with linear growth, 
\begin{equation}\label{e:ss2}
L = x_* t + L_0 + L_1/t + L_2/t^2
\end{equation} 
over the range $\log(m_r/H) > 5.5$, corresponding to $ t \gtrsim 120$. 
The bottom panel of Fig.~\ref{fig:comp_ss} shows the values of $x_*$ and $L_0$ for this case. 
We also show the fits as dotted lines in Fig.~\ref{fig:comp_log}, along with an extrapolation to $\ln(m_rH) = 10$. 
They are barely distinguishable from the logarithmic growth model as a fit to the data, and show how 
the apparent logarithmic growth turns over to a constant in the standard scaling picture. 

It is also notable that the values of $x_*$ do not vary significantly from the 2-parameter fit. 
Clearly $L_0$ changes, but this is to be expected, as it shares the information 
about the initial transients with $L_{-1}$ and $L_{-2}$.
We have also performed bootstrap fits with $x_*$ as a global fit parameter, again finding results clustering around $x_* \simeq 1$.

The tendency of $\xi$ to increase through most of the simulations is also cited in Ref.~\cite{Gorghetto:2020qws} to 
support of the claim of long-term growth. 
This tendency is clear in our simulations too \cite{Hindmarsh:2019csc,Hindmarsh:2021vih}, and can be understood as a transient as the network approaches scaling from low-density configurations 
(see for example Fig. 9 in Ref.~\cite{Hindmarsh:2021vih}).
It was also shown in \cite{Hindmarsh:2021vih} 
 that the tendency to approach the fixed point from low densities 
could be understood in the framework of the VOS model as the  
result of an initial burst of loop production thinning out the network.

Therefore, both sets of simulations support the standard scaling model, 
with consistent values of the asymptotic string density parameter $\xi_* = 1/x_*^2 \simeq 1$, 
which is stable between different initial string densities and to the number of parameters in the fits used to extract it. 
The value can be understood as approximately one Hubble-sized loop being produced per Hubble time 
per Hubble volume, and subsequently decaying in about a Hubble time \cite{Saurabh:2020pqe}.  
The VOS model gives us the physical understanding to confidently extrapolate the result.  

On the other hand, the alternative long-term logarithmic growth model presented in Refs.~\cite{Gorghetto:2018myk,Gorghetto:2020qws} 
lacks a dynamical framework which justifies the logarithmic fits, or their extrapolation. 
Appealing to the excellence of the fits is not enough, as any smooth function over a finite interval 
can be arbitrarily well approximated by an expansion in a set of basis functions. 
The method for extracting the coefficient of the logarithm 
ignores its sensitivity to the initial string density. 
The instability of the coefficient of the logarithm between different initial string densities and 
different methods for extracting the parameter 
is a sign that the model is not 
the correct description of the long-term behaviour. 

It was also pointed out in Ref.~\cite{Hindmarsh:2021vih} that there is a potentially decisive test 
between the two scenarios: whether simulations give asymptotic values 
of $\xi$ significantly larger than O(1) or not.  Simulations of the U(1) complex field models to date 
have final values of $\xi\simeq1$, including the ones presented in Ref.~\cite{Gorghetto:2020qws} as we have established here. 
The slowly-growing $\xi$ evolution shown in the simulations 
always has $\xi \lesssim 1$ at late times, as is consistent with a transient 
bringing the the system to the standard constant-$\xi$ scaling 
solution with $\xi \simeq 1$.  

In summary, the standard scaling scenario with $\xi \simeq 1$ is consistent with all simulations to date, and 
the proposal in Refs.~\cite{Gorghetto:2018myk,Gorghetto:2020qws} to replace the standard scaling scenario by the long-term logarithmic growth model is unjustified. 
Observational predictions based on long-term logarithmic growth 
\cite{Gorghetto:2018myk,Gorghetto:2020qws,Gorghetto:2021fsn} are therefore unsubstantiated.

\emph{Note added:} 
Recently, another group \cite{Buschmann:2021sdq} has used adaptive mesh refinement to simulate axion strings to $\log(m_r/H) \simeq 9$.  The initial string density of their only simulation 
is close to that of the ``preferred'' simulation set of Ref.~\cite{Gorghetto:2020qws}.  
We have checked that the data for $\xi$ in Ref.~\cite{Buschmann:2021sdq} are consistent with an approach to standard scaling with $\xi_* = 1.31 \pm 0.05$, in accord with our results. Therefore, although the data is presented within the framework of the logarithmic growth model, it supports standard scaling.  

\emph{Second note added:} 
More recently, another group also analysed digitised data from Ref.~\cite{Gorghetto:2020qws}, presenting the results in \cite{Hoof:2021jft}. 
They also noted the strong anticorrelation between the global fit parameters $c_1$ and $c_0$. 
The error on $c_1$ quoted there does not include the uncertainty arising from the choice of initial string density, as 
the samples chosen for the bootstrap are taken from the complete set of digitised points. 
We thank the authors for enlightening exchanges over the fitting methods employed in Ref.~\cite{Gorghetto:2020qws}, 
which has enabled us to correct an error of interpretation and to improve the discussion.

\bibliographystyle{JHEP}
\bibliography{axion} % Produces the bibliography via BibTeX.

\end{document}